\documentclass[aps,nopacs,nokeys,superscriptaddress,11pt,twoside]{revtex4}

\usepackage{graphicx,epic,eepic,epsfig,amsmath,latexsym,amssymb,verbatim,revsymb}

\usepackage{color}
\usepackage{theorem}

\def\squareforqed{\hbox{\rlap{$\sqcap$}$\sqcup$}}
\def\qed{\ifmmode\squareforqed\else{\unskip\nobreak\hfil
\penalty50\hskip1em\null\nobreak\hfil\squareforqed
\parfillskip=0pt\finalhyphendemerits=0\endgraf}\fi}
\def\endenv{\ifmmode\;\else{\unskip\nobreak\hfil
\penalty50\hskip1em\null\nobreak\hfil\;
\parfillskip=0pt\finalhyphendemerits=0\endgraf}\fi}

\mathchardef\ordinarycolon\mathcode`\:
\mathcode`\:=\string"8000
\def\vcentcolon{\mathrel{\mathop\ordinarycolon}}
\begingroup \catcode`\:=\active
  \lowercase{\endgroup
  \let :\vcentcolon
  }

\newcommand{\nc}{\newcommand}
\nc{\rnc}{\renewcommand}
\nc{\beq}{\begin{equation}}
\nc{\eeq}{{\end{equation}}}
\nc{\beqa}{\begin{eqnarray}}
\nc{\eeqa}{\end{eqnarray}}
\nc{\lbar}[1]{\overline{#1}}
\nc{\bra}[1]{\langle#1|}
\nc{\ket}[1]{|#1\rangle}
\nc{\ketbra}[2]{|#1\rangle\!\langle#2|}
\nc{\braket}[2]{\langle#1|#2\rangle}
\nc{\proj}[1]{| #1\rangle\!\langle #1 |}
\nc{\avg}[1]{\langle#1\rangle}
\nc{\Rank}{\operatorname{Rank}}
\nc{\smfrac}[2]{\mbox{$\frac{#1}{#2}$}}
\nc{\tr}{\operatorname{Tr}}
\nc{\ox}{\otimes}
\nc{\dg}{\dagger}
\nc{\dn}{\downarrow}
\nc{\cA}{{\cal A}}
\nc{\cB}{{\cal B}}
\nc{\cC}{{\cal C}}
\nc{\cD}{{\cal D}}
\nc{\cE}{{\cal E}}
\nc{\cF}{{\cal F}}
\nc{\cG}{{\cal G}}
\nc{\cH}{{\cal H}}
\nc{\cI}{{\cal I}}
\nc{\cJ}{{\cal J}}
\nc{\cK}{{\cal K}}
\nc{\cL}{{\cal L}}
\nc{\cM}{{\cal M}}
\nc{\cN}{{\cal N}}
\nc{\cO}{{\cal O}}
\nc{\cP}{{\cal P}}
\nc{\cR}{{\cal R}}
\nc{\cS}{{\cal S}}
\nc{\cT}{{\cal T}}
\nc{\cX}{{\cal X}}
\nc{\cZ}{{\cal Z}}
\nc{\csupp}{{\operatorname{csupp}}}
\nc{\qsupp}{{\operatorname{qsupp}}}
\nc{\var}{\operatorname{var}}
\nc{\rar}{\rightarrow}
\nc{\lrar}{\longrightarrow}
\nc{\polylog}{\operatorname{polylog}}
\nc{\1}{{\openone}}
\nc{\sign}{{\operatorname{sign}}}

\def\a{\alpha}
\def\b{\beta}

\nc{\RR}{{{\mathbb R}}}
\nc{\CC}{{{\mathbb C}}}
\nc{\FF}{{{\mathbb F}}}
\nc{\NN}{{{\mathbb N}}}
\nc{\ZZ}{{{\mathbb Z}}}
\nc{\PP}{{{\mathbb P}}}
\nc{\QQ}{{{\mathbb Q}}}
\nc{\UU}{{{\mathbb U}}}
\nc{\EE}{{{\mathbb E}}}
\nc{\id}{{\operatorname{id}}}

\nc{\be}{\begin{equation}}
\nc{\ee}{{\end{equation}}}
\nc{\bea}{\begin{eqnarray}}
\nc{\eea}{\end{eqnarray}}
\nc{\<}{\langle}
\rnc{\>}{\rangle}
\nc{\Hom}[2]{\mbox{Hom}(\CC^{#1},\CC^{#2})}
\nc{\rU}{\mbox{U}}

\nc{\ob}[1]{#1}

\begin{document}

\title{The maximum output $\mathbf{p}$-norm of quantum channels\protect\\
       is not multiplicative for any $\mathbf{p>2}$}

\author{Andreas Winter}
\affiliation{Department of Mathematics, University of Bristol, Bristol BS8 1TW, U.K.}
\affiliation{Quantum Information Technology Lab, National University of Singapore,
 2 Science Drive 3, Singapore 117542}
\email{a.j.winter@bris.ac.uk}

\date{30 July 2007}

\maketitle

\noindent
{\bf Introduction and context.}
In quantum information theory, just as in its classical counterpart,
operational capacities (of information transmission over channels,
of state distillation or preparation procedures, and the like)
are most naturally expressed in terms of (von Neumann) entropies
$S(\rho) = -\tr\rho\log\rho$~\cite{BS}. Usually the
formulas involve optimisation of the entropic quantity in question
over finitely many parameters; examples include
the \emph{entanglement of formation} of a bipartite state~\cite{BDSW},
the so-called \emph{Holevo capacity} of a quantum channel~\cite{Holevo},
or the \emph{minimum output entropy} of a channel~\cite{conjecture}.

Since we are dealing with an asymptotic theory of information -- in the
simplest case, this means looking at many independent realisations of the
state or channel under considerations -- the natural and imminently
important question arises, if these quantities are extensive; in
information theory, this is the additivity problem, which asks if
quantities like entanglement of formation, Holevo capacity, minimum
output entropy, etc.~are additive under tensor products. While all these
conjectures remain open to date, interestingly, Shor~\cite{Shor} has shown that
the three mentioned are actually equivalent.
See~\cite{Problem10} for a general exposition, and pointers to the literature.
Here, we will deal with a related conjecture, on minimum output
\emph{R\'{e}nyi} entropies, and exhibit counterexamples for all
R\'{e}nyi parameters $p>2$.

\bigskip\noindent
{\bf The conjecture.}
For quantum channels ${\cal N}$ (i.e., completely positive, trace
preserving [cptp] maps between -- finite -- quantum
systems) one considers the maximum output $p$-norm ($p>1$),
\begin{equation}
  \label{eq:p-norm}
  \nu_p({\cal N}) = \max_\rho \| {\cal N}(\rho) \|_p,
\end{equation}
where the maximum is over all normalised positive semidefinite density
operators, and $\| X \|_p = \left( \tr |X|^p \right)^{1/p}$ is the
operator $p$-norm. W.l.o.g.~the maximisation may be restricted to pure
states $\rho = \proj{\psi}$.
In conjunction with the conjectured additivity of the minimum output
entropy of the channel, it has been conjectured that $\nu_p$ is
\emph{multiplicative} for $p>1$~\cite{conjecture}:
\begin{equation}
  \label{eq:mult}
  \nu_p\bigl( {\cal N}_1 \ox {\cal N}_2 \bigr) = \nu_p({\cal N}_1) \nu_p({\cal N}_2).
\end{equation}
Indeed, this multiplicativity for $p$ sufficiently close to $1$ would imply 
the additivity of the minimum output entropy of the channel, and hence~\cite{Shor}
the other ``standard'' additivity conjectures of quantum information theory.
An easy way of making this link is to consider R\'{e}nyi $p$-entropy,
$S_p(\rho) = \frac{1}{1-p}\log\tr\rho^p$, instead of the $p$-purity
$\tr\rho^p$. Clearly then, the minimum output $p$-entropy of the
channel is $\frac{p}{1-p}\log\nu_p({\cal N})$, turning multiplicativity
into additivity. To finish the argument, all that is needed is the
observation that $\lim_{p\rightarrow 1} S_p(\rho) = S(\rho)$.

From the elementary multiplicativity of the $p$-norm itself, the
inequality ``$\geq$'' in (\ref{eq:mult}) is trivial, so the question is
always if ``$\leq$'' holds. Now, Holevo and Werner~\cite{HolevoWerner} have shown
that (\ref{eq:mult}) cannot hold universally: there is a channel
${\cal N}_{\rm HW}$ which provides a counterexample for $p > 4.79$, in the
sense that for all such $p$,
\[
  \nu_p\bigl( {\cal N}_{\rm HW} \ox {\cal N}_{\rm HW} \bigr) > \nu_p({\cal N}_{\rm HW})^2.
\]

On the other hand, King and Ruskai have argued~\cite{conjecture-2}
that (\ref{eq:mult}) should still hold true for $1<p \leq 2$. Incidentally,
for the Holevo-Werner channel ${\cal N}_{\rm HW}$, and a whole class
containing it, multiplicativity has indeed been shown for
$1 < p \leq 2$ and arbitrary number of tensor factors~\cite{AlickiFannes}.
This poses the natural problem of constructing counterexamples to (\ref{eq:mult})
for all $p>2$. In the rest of the paper,
we show that approximately randomising channels~\cite{rand} provide exactly that.

\bigskip\noindent
{\bf Random unitary channels.}
These are channels of the form
\begin{equation}
  \label{eq:rand}
  {\cal N}:\rho \longmapsto \frac{1}{n} \sum_{i=1}^n V_i \rho V_i^\dagger,
\end{equation}
with unitaries $V_i$ of the underlying $d$-dimensional Hilbert space
(More generally, one could allow variable probability weights
for different $V_i$, but we won't need that here.)

Following~\cite{rand}, we call ${\cal N}$ \emph{$\epsilon$-randomising}, if 
\begin{equation}
  \label{eq:opnorm-epsrand}
  \forall\rho\qquad  \left\| {\cal N}(\rho) - \frac{1}{d}\1 \right\|_\infty
                                                   \leq \frac{\epsilon}{d}.
\end{equation}
There, it is shown that for $0<\epsilon<1$, $\epsilon$-randomising channels
exist in all dimensions $d > \frac{10}{\epsilon}$,
with $n = \frac{134}{\epsilon^2} d \log d$ -- in fact, randomly
picking the $V_i$ from the Haar measure on the unitary group
will, with high probability, yield such a channel.

\bigskip\noindent
{\bf Lemma 1.} For a random unitary channel ${\cal N}$ and its
complex conjugate,
$\overline{\cal N}:\rho \mapsto \frac{1}{n} \sum \overline{V_i} \rho \overline{V_i}^\dagger$,
one has $\nu_p({\cal N}\ox\overline{\cal N}) \geq \frac{1}{n}$.

\medskip\noindent
\emph{Proof.}
We use the maximally entangled state
$\Phi_d = \frac{1}{d}\sum_{\a\b} \ket{\a\a}\!\bra{\b\b}$ as test state:
\[\begin{split}
  \nu_p({\cal N}\ox\overline{\cal N})
      &\geq \left\| ({\cal N}\ox\overline{\cal N})\Phi_d \right\|_p \\
      &=    \left\| \frac{1}{n^2}\sum_{ij=1}^n (V_i\ox\overline{V_j}) 
                                                 \Phi_d 
                                               (V_i\ox\overline{V_j})^\dagger \right\|_p \\
      &=    \left\| \frac{1}{n}\Phi_d
                     + \frac{1}{n^2}\sum_{i\neq j} (V_i\ox\overline{V_j}) 
                                                     \Phi_d 
                                                   (V_i\ox\overline{V_j})^\dagger \right\|_p
       \geq \frac{1}{n},
\end{split}\]
where in the third line we have invoked the $U\ox\overline{U}$-invariance
of $\Phi_d$, for all $n$ occurrences of $i=j$. For the final inequality,
observe that the largest eigenvalue $\lambda_1$ of
$\omega := ({\cal N}\ox\overline{\cal N})\Phi_d$ is $\geq \frac{1}{n}$,
and denoting the other eigenvalues $\lambda_{\a}$,
$\| \omega \|_p = \left( \sum_{\a} \lambda_{\a}^p \right)^{1/p} \geq \lambda_1$,
and we are done.
\qed

\bigskip\noindent
{\bf Lemma 2.} If the channel ${\cal N}$ is $\epsilon$-randomising,
then, with $p>1$,
\[
  \nu_p({\cal N}) = \nu_p(\overline{\cal N}) \leq \left( \frac{1+\epsilon}{d} \right)^{1-1/p}.
\]

\medskip\noindent
\emph{Proof.}
Clearly, ${\cal N}$ and $\overline{\cal N}$ have the same maximum
output $p$-norm. For the former, observe that the $\epsilon$-randomising
condition implies, for arbitrary input state $\rho$,
$\| {\cal N}(\rho) \|_\infty \leq \frac{1+\epsilon}{d}$, in other
words, all eigenvalues $\lambda_{\a}$ of the output state ${\cal N}(\rho)$
are bounded between $0$ and $\frac{1+\epsilon}{d}$, besides summing to $1$. 

Hence, by convexity of the function $x\mapsto x^p$, the $p$-norm
$\| {\cal N}(\rho) \|_p = \left( \sum_{\a} \lambda_{\a}^p \right)^{1/p}$
is maximised, under these constraints, at a spectrum with
largest eigenvalue $\frac{1+\epsilon}{d}$, with multiplicity
$\frac{d}{1+\epsilon}$, and the remaining eigenvalues $0$,
yielding 
$\| {\cal N}(\rho) \|_p =    \left( \sum_{\a} \lambda_{\a}^p \right)^{1/p}
                        \leq \left( \frac{d}{1+\epsilon}
                                    \left( \frac{1+\epsilon}{d} \right)^p \right)^{1/p}
                        =    \left( \frac{1+\epsilon}{d} \right)^{1-1/p}.$
\qed

\bigskip\noindent
{\bf Main result.} Now fix any $0<\epsilon<1$, and a family
of $\epsilon$-randomising maps ${\cal N}$ for all sufficiently
large dimensions $d$ and $n = O(d \log d)$ as above. Then,
for any $p>2$ and sufficiently large $d$,
\begin{equation}
  \label{eq:main}
  \nu_p({\cal N}) \nu_p(\overline{\cal N})
                      \leq \left( \frac{1+\epsilon}{d} \right)^{2-2/p}
                      \ll  \frac{1}{n}
                      \leq \nu_p({\cal N}\ox\overline{\cal N}),
\end{equation}
by Lemmas 1 and 2, and since $2-2/p > 1$.
I.e., for this family of channels, the maximum output $p$-norm
is strictly supermultiplicative, eventually.
\qed

\bigskip\noindent
{\bf Discussion.} 
The counterexamples to the multiplicativity of the output $p$-norm
for $p>2$ provided here are interesting in that they are
random unitary channels, which are among the simplest
truly quantum maps -- in fact, the first proofs of multiplicativity
for unital qubit channels~\cite{King1} and depolarising channels~\cite{King2}
relied on this kind of channel structure.
Indeed, unital qubit channels are always random unitary channels
(that is our case with $d=2$), and King~\cite{King1} shows multiplicativity
for such channels at all $p>1$ -- there is no conflict with our
result here, as the bound on $n$ becomes better than $d^2$ only 
for rather large dimension $d$.

We observe, furthermore, that $p=2$ is indeed the limit of
validity of the counterexample(s), since $n \geq d$ for
any $\epsilon$-randomising map. 

Note, however, that in the condition of $\epsilon$-randomisation,
it is not so crucial to have $\epsilon$ small: looking at the
above argument, we see that indeed any constant, or even
any mildly (say, polylogarithmically), in $d$, growing $\epsilon$
will do. Still, it seems that we have to rely on the
strong randomisation via Haar measure unitaries from~\cite{rand}:
all other, more or less explicit, constructions (by Ambainis
and Smith~\cite{AmbainisSmith} or via iterated quantum
expander maps~\cite{q-expanders,Hastings}) only give us
bounds in the $2$-norm, which do imply bounds on the
output $p$-norm but they are too weak for the present purpose.
As a consequence, we don't have any \emph{explicit} counterexamples,
but really only a proof of their existence -- it remains as an
open problem to ``derandomise'' our argument.


\bigskip\noindent
{\bf Note added.} Since the posting of this work on the arXiv,
Patrick Hayden (arXiv:0707.3291) has shown how to extend the range
of the counterexamples to all $p>1$. Interestingly, as $p$ gets
closer to $1$, the dimension of the output has to grow
to infinity for a violation to occur, so the original additivity
conjecture (at $p=1$) is still left intact. This dramatic
finding now focusses the attention on the question of the
multiplicativity of the \emph{minimum} output $p$-norm for
$0 \leq p \leq 1$.

\bigskip\noindent
{\bf Acknowledgments.}
The author is supported by the U.K.~EPSRC (``QIP IRC'' and
an Advanced Research Fellowship) and the EU project ``QAP''.
Thanks to Patrick Hayden, Debbie Leung and Beth Ruskai for discussions
on the multiplicativity conjecture and possible counterexamples,
starting at the BIRS workshop ``Operator Structures in Quantum Information'',
February 2007, and to Toby Cubitt and Ashley Montanaro for
conversations about multiplicativity at $p=0$.
%


\begin{thebibliography}{9}
  \bibitem{AlickiFannes} R. Alicki, M. Fannes, ``Note on Multiple Additivity of
    Minimal Renyi Entropy Output of the Werner-Holevo Channels'',
    Open Systems Inf. Dyn {\bf 11}(4):339-342 (2004);
    arXiv:quant-ph/0407033.

  \bibitem{AmbainisSmith} A. Ambainis, A. Smith, ``Small pseudo-random families
    of matrices: Derandomizing approximate quantum encryption'', in: Proc.
    RANDOM, Springer LNCS 3122, pp.~249-260 (2004);
    arXiv:quant-ph/0404075.
    
  \bibitem{conjecture} G. G. Amosov, A. S. Holevo, R. F. Werner,
    ``On some additivity problems in quantum information theory'',
    arXiv:math-ph/0003002 (2000).

  \bibitem{q-expanders} A. Ben-Aroya, A. Ta-Shma, ``Quantum expanders and the 
    quantum entropy difference problem'',
    arXiv:quant-ph/0702129 (2007).

  \bibitem{BDSW} C. H. Bennett, D. P. DiVincenzo, J. A. Smolin, W. K. Wootters,
    ``Mixed-state entanglement and quantum error correction '',
    Phys. Rev. A {\bf 54}(5):3824-3851 (1996).

  \bibitem{BS} C. H. Bennett, P. W. Shor, ``Quantum Channel Capacities'',
    Science {\bf 303}:1784-1786 (2004).

  \bibitem{Hastings} M. B. Hastings, ``Random Unitaries Give Quantum Expanders'',
    arXiv[quant-ph]:0706.0556 (2007).
    
  \bibitem{rand} P. Hayden, D. Leung, P. W. Shor, A. Winter, ``Randomizing Quantum
    States: Constructions and Applications'', 
    Comm. Math. Phys. {\bf 250}(2):371-391 (2004);
    arXiv:quant-ph/0307104.

  \bibitem{Holevo} A. S. Holevo, ``The capacity of the quantum channel with general signal
    states'', IEEE Trans. Inf. Theory {\bf 44}(1):269-273 (1998);
    arXiv:quant-ph/9611023.
    B. Schumacher, M. D. Westmoreland, ``Sending classical information via noisy quantum
    channels'', Phys. Rev. A {\bf 56}(1):131-138 (1997).

  \bibitem{Problem10} A. S. Holevo, ``Additivity of classical capacity and related problems'',
    http://www.imaph.tu-bs.de/qi/problems/10.html.

  \bibitem{HolevoWerner} A. S. Holevo, R. F. Werner, ``Counterexample to an
    additivity conjecture for output purity of quantum channels'',
    J. Math. Phys. {\bf 43}(9):4353-4357 (2002); 
    arXiv:quant-ph/0203003.

  \bibitem{King1} C. King, ``Additivity for unital qubit channels'', J. Math. Phys.
    {\bf 43}:4641-4653 (2002);
    arXiv:quant-ph/0103156.

  \bibitem{King2} C. King, ``The capacity of the quantum depolarizing channel'',
    IEEE Trans. Inf. Theory {\bf 49}:221-229 (2003);
    arXiv:quant-ph/0204172.

  \bibitem{conjecture-2} C. King, M.-B. Ruskai, ``Comments on multiplicativity 
    of maximal p-norms when p=2'', Quant. Inf. Comput. {\bf 4}(6\&{}7):500-512 (2004);
    arXiv:quant-ph/0401026.


  \bibitem{Shor} P. W. Shor, ``Equivalence of Additivity Questions in Quantum
    Information Theory'',
    Comm. Math. Phys. {\bf 246}(3):453-472 (2004);
    arXiv:quant-ph/0305035.

\end{thebibliography}
\end{document}